\documentclass[a4paper,fleqn,usenatbib]{mnras}

\usepackage{newtxtext,newtxmath}

\usepackage[T1]{fontenc}
\usepackage{ae,aecompl}

\usepackage[capitalise]{cleveref}

\usepackage{graphicx}	
\usepackage{amsmath}	
\usepackage{amssymb}	

\def\ms{\,m\,s$^{-1}$}         
\def\kms{\,km\,s$^{-1}$}       
\def\msol{$M_\odot$}		
\def\rsol{$R_\odot$}		
\def\mstar{$M_*$}		
\def\rstar{$R_*$}		
\def\densstar{$\rho_*$}		
\def\mplanet{$M_{\rm P}$}	
\def\rplanet{$R_{\rm P}$}	
\def\mjup{$M_{\rm Jup}$}	
\def\rjup{$R_{\rm Jup}$}	
\def\teql{$T_{\rm eql}$}
\def\teff{$T_{\rm eff}$}

\def\logg{$\log g_*$}

\def\vsini{$v_* \sin i_*$}

\def\vsinidt{$v_* \sin i_{\rm *,DT}$}

\def\kms{km\, s$^{-1}$}

\def\svsicos{$\sqrt{v_* \sin i_*} \cos \lambda$}

\def\mcmch{MCMC-DT$_{\rm HAR}$} 
\def\mcmcc{MCMC-DT$_{\rm COR}$} 
\def\mcmcrm{MCMC-RM} 

\newcommand{\leftcell}[1]{\multicolumn{1}{l}{#1}}

\newcommand{\rhostar}{\mbox{${\rho}_{\star}$}}

\title[WASP-189b: an ultra-hot Jupiter transiting HR\,5599]{WASP-189b: an ultra-hot Jupiter transiting the bright A star HR\,5599 in a polar orbit\thanks{Based on observations collected at the European Organisation for Astronomical Research in the Southern Hemisphere under ESO programme 0100.C-0847.}}

\author[D.~R.~Anderson et al.]{D.~R.~Anderson,$^{1}$\thanks{E-mail: d.r.anderson@keele.ac.uk}
	   L.~Y.~Temple,$^{1}$
	   L.~D.~Nielsen,$^{2}$ 
	   A.~Burdanov,$^{3}$ 
	   C.~Hellier,$^{1}$\newauthor
	   F.~Bouchy,$^{2}$
	   D.~J.~A.~Brown,$^{4,5}$
	   A.~Collier~Cameron,$^{6}$
	   M.~Gillon,$^{3}$
	   E.~Jehin,$^{3}$\newauthor
	   P.~F.~L.~Maxted,$^{1}$
	   F.~Pepe,$^{2}$
	   D.~Pollacco,$^{4,5}$
	   F.~J.~Pozuelos,$^{3}$
	   D.~Queloz,$^{2,7}$\newauthor
	   D.~S\'egransan,$^{2}$
	   B.~Smalley,$^{1}$
	   A.~H.~M.~J.~Triaud,$^{8}$
	   O.~D.~Turner,$^{2}$
	   S.~Udry$^{2}$\newauthor
	   and
	   R.~G.~West$^{4,5}$
\\
\\
$^{1}$Astrophysics Group, Keele University, Staffordshire ST5 5BG, UK\\
$^{2}$Observatoire de Gen\`eve, Universit\'e de Gen\`eve, 51 Chemin 
  des Maillettes, 1290 Sauverny, Switzerland\\
$^{3}$Space sciences, Technologies and Astrophysics Research (STAR) Institute,  Universit\'e de 
  Li\`ege, Li\`ege 1, Belgium\\
$^{4}$Department of Physics, University of Warwick, Coventry CV4 7AL, UK\\
$^{5}$Centre for Exoplanets and Habitability, University of Warwick, Gibbet Hill Road, Coventry CV4 7AL, UK\\
$^{6}$SUPA, School of Physics and Astronomy, University of St. Andrews, 
  North Haugh, Fife KY16 9SS, UK\\
$^{7}$Cavendish Laboratory, J J Thomson Avenue, Cambridge CB3 0HE, UK\\
$^{8}$School of Physics \& Astronomy, University of Birmingham, Edgbaston, Birmingham, B15 2TT, UK\\
}

\date{Accepted XXX. Received YYY; in original form 2018 September 13}

\pubyear{2018}

\begin{document}
\label{firstpage}
\pagerange{\pageref{firstpage}--\pageref{lastpage}}
\maketitle

\begin{abstract}
We report the discovery of WASP-189b: an ultra-hot Jupiter in a 2.72-d transiting orbit around the $V = 6.6$ A star WASP-189 (HR\,5599). 
We detected periodic dimmings in the star's lightcurve, first with the WASP-South survey facility then with the TRAPPIST-South telescope. We confirmed that a planet is the cause of those dimmings via line-profile tomography and radial-velocity measurements using the HARPS and CORALIE spectrographs. Those reveal WASP-189b to be an ultra-hot Jupiter (\mplanet\ = 2.13 $\pm$ 0.28\,\mjup; \rplanet\ = 1.374 $\pm$ 0.082\,\rjup) in a polar orbit ($\lambda = 89.3 \pm 1.4^\circ$; $\Psi = 90.0 \pm 5.8^\circ$) around a rapidly rotating A6IV--V star (\teff\ = 8000 $\pm$ 100\,K; \vsini\ $\approx$ 100\,\kms). 
We calculate a predicted equilibrium temperature of \teql\ = 2641 $\pm$ 34\,K, assuming zero albedo and efficient redistribution, which is the third hottest for the known exoplanets. 
WASP-189 is the brightest known host of a transiting hot Jupiter and the third-brightest known host of any transiting exoplanet. We note that of the eight hot-Jupiter systems with \teff\ $>$ 7000\,K, seven have strongly misaligned orbits, and two of the three systems with  \teff\ $\geq$ 8000\,K have polar orbits (the third is aligned).
\end{abstract}




\section{Introduction}
With $\sim$750 transiting planets well studied the emphasis is on pushing planet discovery into regimes where there are few known planets. 
Of the 580 host stars listed in the TEPCat database \citep{2011MNRAS.417.2166S}, only 9 have stellar effective temperature \teff $>$ 7000\,K.
This is because the spectral lines of very hot, rapidly rotating stars are sparse and broad, which complicates the radial-velocity verification of planets.
Hence we know little about the formation and evolution of planets around hot stars. 

Under the core accretion model of planet formation, giant planets are expected to be more common around more massive stars \citep{1996Icar..124...62P}, with an occurrence rate predicted by \citet{2008ApJ...673..502K} to increase linearly with stellar mass over the range 0.4--3\,\msol. 
Indeed, \citet{2008PASP..120..531C} found that the occurrence rate of giant planets (planet mass \mplanet\ = 0.3--10\,\mjup; orbital period $P$ = 2--2000\,d) is ten times lower for M\,dwarfs (1 per cent) than for FGK stars (10.5 per cent). 
Whether this trend continues to A stars is unknown.
\citet{2017A&A...599A..57B} reported an occurrence rate of $4^{+10}_{-4}$ per cent for A-type stars (\mplanet\ = 0.1--10\,\mjup; $P$ = 1--1000\,d), which is too imprecise to tell. 

An additional motivation to discover planets around hot stars is their potential for characterisation. 
As hot stars are intrinsically luminous, many bright stars (which are easier to study) are hot stars: of the Bright Star Catalogue \citep{1991bsc..book.....H}, 41 per cent are listed as OBA stars and 21 per cent are listed as A stars.
Also, planets in close orbits around hot stars are intensely irradiated, which facilitates the probing of their atmospheres. 
For example, a great deal of work has been done to characterise the atmosphere of WASP-33b (\citealt{2010MNRAS.407..507C}; \teff\ = 7430\,K; $V = 8.2$; $P = 1.22$\,d) from detections of its thermal emission \citep{2011MNRAS.416.2096S, 2012ApJ...754..106D, 2013A&A...550A..54D, 2015ApJ...806..146H, 2015A&A...584A..75V, 2017AJ....154..221N, 2018AJ....155...83Z}.   

In their initial configuration the WASP survey facilities operated with 200-mm f/1.8 lenses, as described by \citet{2006PASP..118.1407P}, targetting stars in the range $V$ = 9--13. WASP-South ran in this mode from 2006 to 2012, resulting in the discovery of 125 planets (some jointly with SuperWASP-North). But WASP-South saturated at $V$ = 9, and the brightest planet-host found in the South was WASP-18 at $V$ = 9.3 \citep{2009Natur.460.1098H}.  Since several brighter hosts were known in the North we decided to switch to Canon 85-mm f/1.2 lenses, with an SDSS $r$-band filter and shorter exposures (three 20-s exposures per pointing instead of two 30-s exposures), so covering the range $V$ = 6.5--11. WASP-South ran in this configuration from 2012 to 2015. 

This survey mode is similar to that of the KELT project \citep{2007PASP..119..923P}, which uses a Mamiya 80-mm f/1.9 lens, and which has been vindicated by spectacular discoveries such as KELT-9 \citep{2017Natur.546..514G} and KELT-20/MASCARA-2 (\citealt{2017AJ....154..194L,2018A&A...612A..57T}), which are both $V$ = 7.6 systems. KELT-9b, which transits the hottest known hot-Jupiter host (\teff $\approx$ 10\,200\,K), is already proving invaluable for atmospheric characterisation \citep{2018Natur.560..453H}.   

In this paper we present the discovery of WASP-189b, an ultra-hot Jupiter transiting an even brighter star, being in a polar orbit around the mid-A star WASP-189 (HR\,5599; $V=6.6$; \teff\ = 8000\,K).

\section{Photometric and spectroscopic observations}

\begin{figure}
\includegraphics[width=90mm]{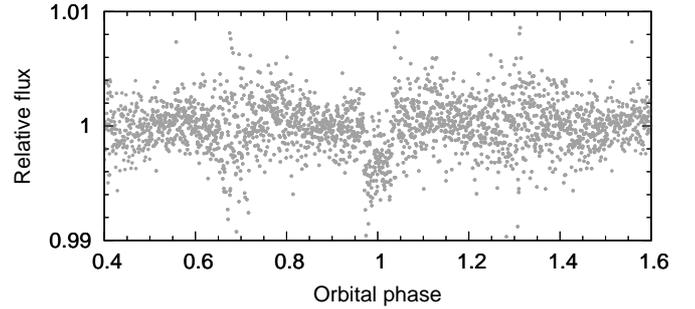}
\caption{WASP-South survey lightcurve folded on the adopted ephemeris from \cref{tab:mcmc} and binned (bin width equivalent to 2\,min). 
\label{fig:wasp}}
\end{figure}

We present in \cref{tab:obs} a summary of our photometric and spectroscopic observations of WASP-189. 
We observed WASP-189 during 2012 Jul to 2013 Aug with the modified WASP-South survey facility.  Our search techniques are described in \citet{2006MNRAS.373..799C, 2007MNRAS.380.1230C}.
We detected a dip of a few mmag and with a periodicity of 2.72\,d in the WASP-South lightcurve (\cref{fig:wasp}). 
Initial radial-velocity (RV) measurements from the CORALIE spectrograph on the Swiss {\it Euler} 1.2-m telescope \citep{2000A&A...354...99Q} indicated motion of a few 100\,\ms, so we added the target to our program searching for planets around hot stars. 

\begin{figure*}
\centering
\includegraphics[width=180mm]{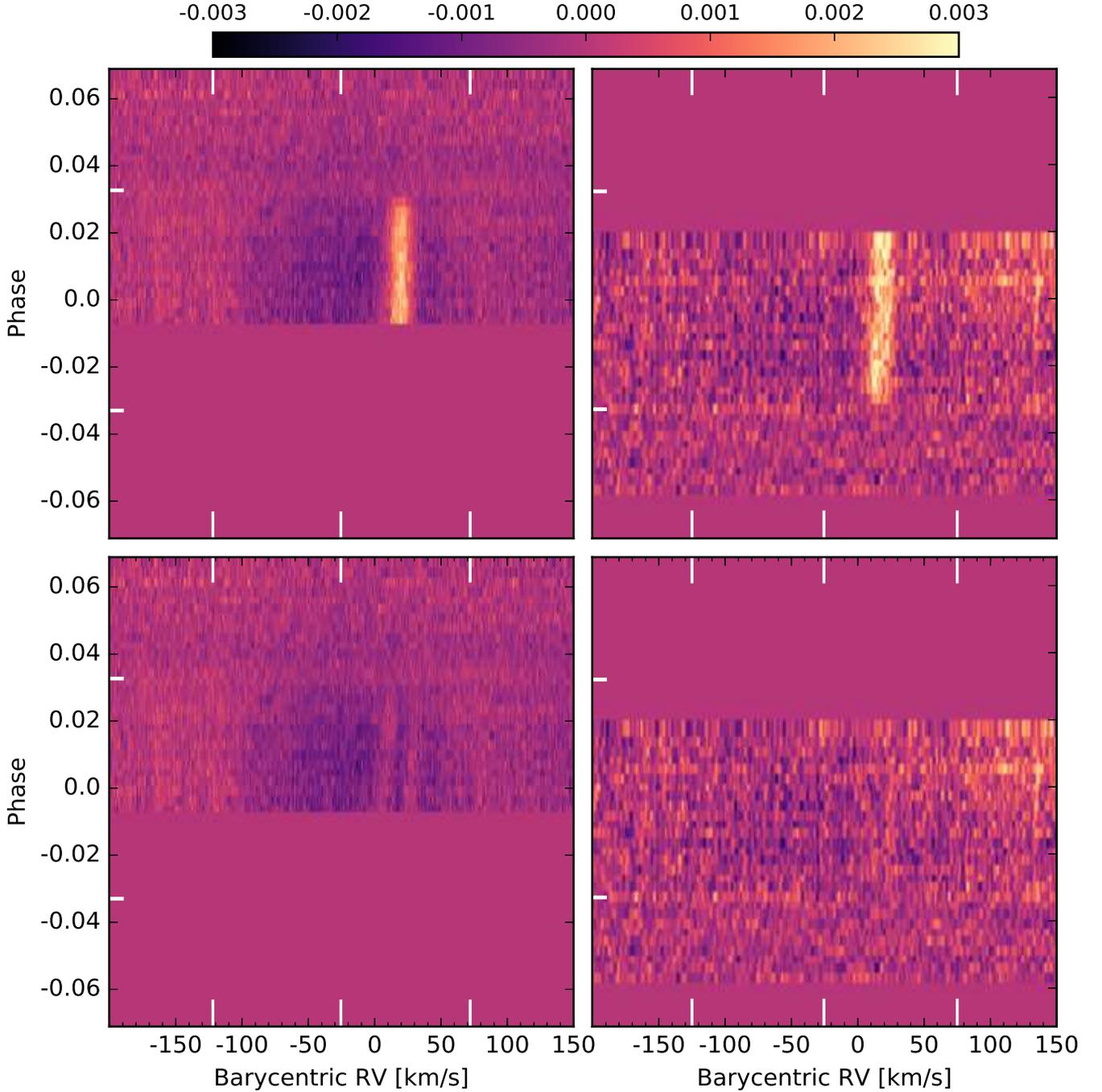}
\caption{
{\it Top row}: 
Doppler tomograms (residual maps of the CCF time-series) through transits of WASP-189b from HARPS (left) and CORALIE (right), phased on the empherides of \cref{tab:mcmc}. 
The mean out-of-transit CCF (which is a CCF made up of the mean value in the out-of-transit portion of the time-series at each wavelength) has been subtracted, leaving the bright signature of the starlight blocked by the planet during transit. 
The stellar velocity of the planet trace appears constant, indicating a polar orbit, and the offset of the trace from the systemic velocity (on the receding limb of the star) indicates a non-zero impact parameter. 
The white, vertical marks indicate the positions of the systemic velocity ($\gamma$) and the limits of the stellar rotation velocity ($\gamma$ $\pm$ \vsinidt).
The white, horizontal marks indicate the start and end of transit (first and fourth contact). 
{\it Bottom row}: 
The same as the top row, but after removal of the best-fitting planetary signature model. 
\label{fig:tomog}}
\end{figure*}

\begin{figure}
\centering
\includegraphics[width=90mm]{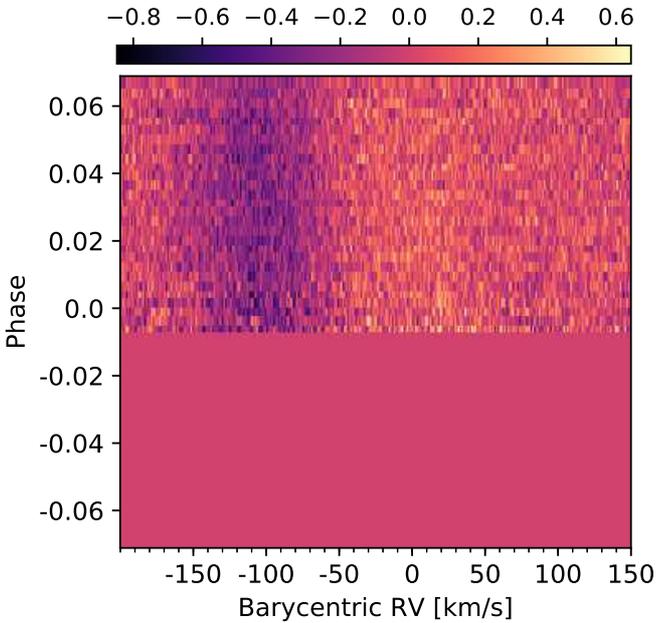}
\caption{
Doppler tomogram obtained from the sky spectra observed by the secondary fibre of HARPS whilst its primary fibre was on target. Due to the absence of a similar trace in the secondary fibre tomogram, we concluded that the trace seen in the primary-fibre tomogram (\cref{fig:tomog}, top-left panel) was caused by WASP-189b rather than the Moon.
The range of the plot is identical to \cref{fig:tomog}. The intensity scale was changed to bring out the detail, as the throughputs of the two fibres differ and no stellar profile was subtracted from the secondary fibre CCFs.
\label{fig:tomog-sky}}
\end{figure}

Using HARPS on the ESO 3.6-m telescope \citep{2000SPIE.4008..582P}, we obtained a series of spectra during the transit night of 2018 Mar 26. 
Using the HARPS Data Reduction Software (DRS), we computed average stellar line profiles, or cross-correlation functions (CCFs), by cross-correlating the HARPS spectra with a weighted A0 binary mask from which we had removed various tellurics and ISM lines \citep{1996A&AS..119..373B,2002Msngr.110....9P}. The time-series of these CCFs constitute the HARPS Doppler tomography data set, DT$_{\rm HAR}$. We discarded the last two spectra of the series as they were visibly affected by twilight. 
In the resulting tomogram (top-left panel of \cref{fig:tomog}), there is a trace indicative of a planet in a polar orbit crossing the red-shifted hemisphere of the star. 
However, due to ephemeris uncertainty (our initial attempts to refine the ephemeris with additional transit photometry were unsuccessful) we had missed the ingress of the event and we were concerned about possible lunar contamination. 
The Moon, which was 77 per cent illuminated and at a distance of $\sim$$95^\circ$ from the target, had set at 06:46 UT. 
The barycentric Earth velocity was 18.5\,\kms. 
The trace in the tomogram occurs at a similar velocity and is apparent in the CCFs taken up to 06:04 UT.
We cross-correlated with the A0 mask the sky spectra taken simultaneously using the secondary HARPS fibre and found no signal (\cref{fig:tomog-sky}), which indicates that the signal in the target tomogram is not due to the Moon. 

Using CORALIE, we obtained a series of spectra during the transit night of 2018 Jul 13.
The Moon posed no issue as it was only 2 per cent illuminated 
and it set at 23:01 UT (26\,min before the CORALIE sequence began). 
We again computed CCFs using the A0 mask to produce the CORALIE Doppler tomography data set, DT$_{\rm COR}$.
In the resulting tomogram (top-right panel of \cref{fig:tomog}), there is a trace indicative of a planet in a polar orbit crossing the red-shifted hemisphere of the star. This is similar to the trace in the DT$_{\rm HARPS}$ tomogram, which occurs at a similar barcyentric RV.

\begin{figure}
\includegraphics[width=90mm]{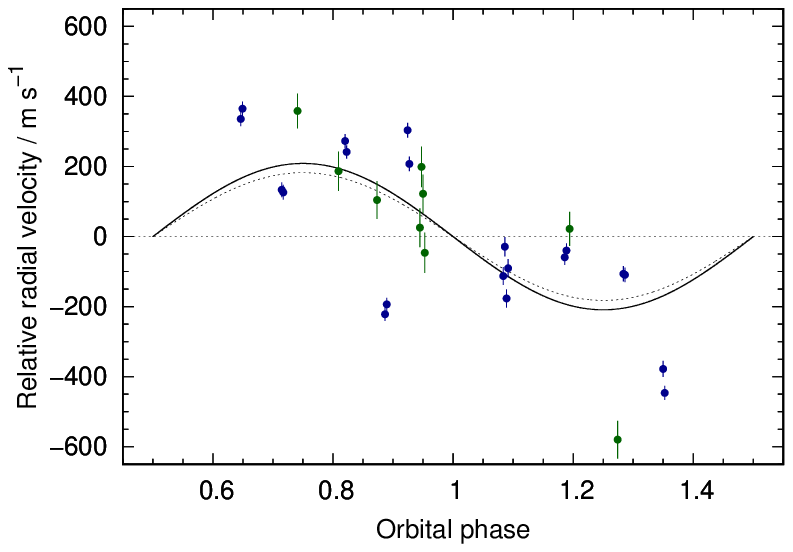}
\caption{Radial-velocities of WASP-189 as extracted using the DRS (green = CORALIE; blue = HARPS). We were motivated by the large scatter to improve the extraction technique. The solid line is the best-fitting orbit from an MCMC including the DRS RVs and all of the photometry. 
The dotted line is the best-fitting orbit to the improved RVs (\cref{fig:rv-tran}). 
\label{fig:rv-drs}}
\end{figure}

\begin{figure}
\includegraphics[width=90mm]{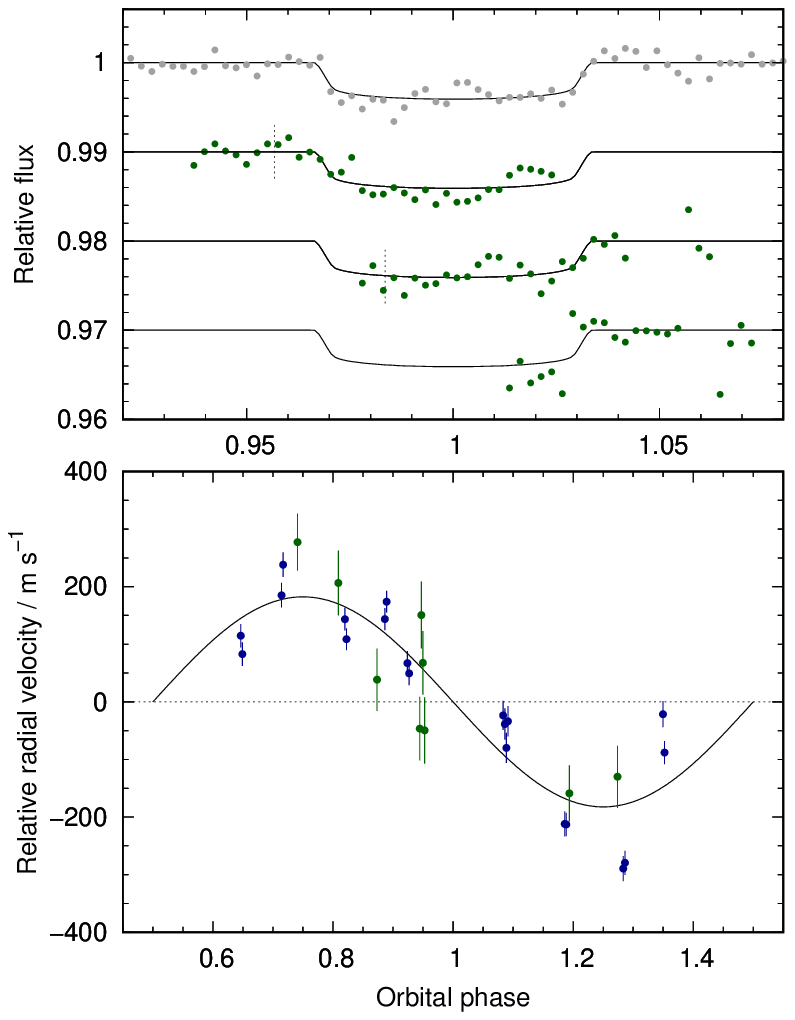}
\caption{{\it Top panel}: Transit photometry from WASP-South (grey) and TRAPPIST-South (green) folded on the adopted ephemeris from \cref{tab:mcmc}, binned (bin width equivalent to 10\,min) and offset for display.
The adopted transit model is superimposed. Vertical dashed lines indicate partitioning due to meridian flips. 
{\it Bottom panel}: Radial velocities from CORALIE (green) and HARPS (blue) with the adopted circular orbital model.
\label{fig:rv-tran}}
\end{figure}

We also obtained spectra around the orbit, 5 with CORALIE and 16 with HARPS. We again computed CCFs by cross-correlating the spectra with the same A0 mask as before. 
We used the DRS to calculate RVs from all of the CCFs (including DT$_{\rm HAR}$ and DT$_{\rm COR}$), which it does by fitting a Gaussian, but we found the scatter about the best-fitting orbit to be large (\cref{fig:rv-drs}; weighted RMS = 163\,\ms). 
So we instead calculated the RVs by fitting the CCFs, which we detrended using a linear function fit to the wings, with a simple rotationally broadened profile \citep{1976oasp.book.....G}:

\begin{equation}
{\rm contrast}(v) = c_1  \sqrt{1 - \left(\frac{(v-RV)}{RV}\right)^2} + c_2 \left(1 - \left(\frac{(v-RV)}{RV}\right)^2\right) + c_3
\end{equation}

\noindent where $v$ is velocity. The first and second terms parameterise rotation and limb-darkening, respectively, and the third term is an offset. 
The scatter about the best-fitting orbit was thus much reduced (bottom panel of \cref{fig:rv-tran}, note the reduced range in the RV axis compared to \cref{fig:rv-drs}; weighted RMS = 70\,\ms).

We conducted photometric follow-up using the 0.6-m TRAPPIST-South imager \citep{2011EPJWC..1106002G,2011Msngr.145....2J} and a red continuum cometary narrowband filter on the transit nights of 2018 Jul 13, 2018 Jul 24 and 2018 Aug 4. We processed the resulting time-series of images to obtain lightcurves using standard differential photometry techniques. 
In the lightcurve from 2018 Jul 13 there is an ingress coincident with the start of the planetary signature in the CORALIE tomogram, which was taken simultaneously (\cref{fig:tomog,fig:rv-tran}). 
When phased on our adopted transit ephemeris (\cref{sec:mcmc}), the egresses in the TRAPPIST-South lightcurves are concident with the end of the planetary signature in the HARPS tomogram (\cref{fig:tomog,fig:rv-tran}).

\begin{table*}
\small
\centering
\small
\caption{Summary of observations}
\label{tab:obs}
\begin{tabular}{lllllll}
\hline
\hline
\leftcell{Facility} & \leftcell{Start time} & \leftcell{Culmination} & \leftcell{End time} & \leftcell{$N_{\rm exp}$} & \leftcell{$T_{\rm exp}$} & \leftcell{Notes$^a$}\\
 & \leftcell{[airmass]} & \leftcell{[airmass]} & \leftcell{[airmass]} & & (s) & \\
\hline
\\
{\it Photometry}\\
WASP-South						& 2012 Jul 03				& \ldots		& 2013 Aug 10	& 48\,307	& 20	& filter = 400--700 nm\\
TRAPPIST-South					& 2018 Jul 13 23:09 [1.17]	& 00:19 [1.11]	& 04:59 [3.10]	& 1\,247	& 10	& filter = rcnb; MF = 8313.516 \\
TRAPPIST-South					& 2018 Jul 24 23:12 [1.12]	& 23:36 [1.11]	& 03:39 [2.22]	& 969		& 10	& filter = rcnb; MF = 8324.485 \\
TRAPPIST-South					& 2018 Aug 04 23:09 [1.12]	& \ldots & 03:10 [2.49]	& 780		& 10	& filter = rcnb\\
\\
{\it Spectroscopy}\\
Swiss/CORALIE				& 2017 May 19 				& \ldots		& 2018 May 08 	& 5		& 900--1400	& orbit \\
ESO3.6/HARPS				& 2018 Mar 20				& \ldots		& 2018 Mar 25 	& 16	& 600	& orbit\\
ESO3.6/HARPS				& 2018 Mar 26 03:37 [2.04]	& 07:28 [1.11]	& 10:27 [1.55] 	& 39	& 600	& transit; Sun alt. = $-12^\circ$ at 09:59\\
Swiss/CORALIE				& 2018 Jul 13 23:27 [1.14]	& 00:19 [1.11]	& 04:11 [2.05] 	& 28	& 600	& transit \\
\\
\hline
\end{tabular}
{\it Note:} 
\raggedright
$^a$  Filter `rcnb' = red continuum narrowband.
`MF' indicates a meridian flip occurred at the noted time (BJD $-$ 2450000).
\end{table*}

\section{Stellar properties of WASP-189}
\label{sec:stellar}

We co-added the individual HARPS spectra from the transit night to obtain an average signal-to-noise of $\sim$200:1. 
We estimated the stellar effective temperature, \teff\ = 8000 $\pm$ 100\,K, from the Balmer lines, assuming a stellar surface gravity of \logg\ = 4.0. We estimated the projected stellar rotation speed \vsini\ = $100 \pm 5$\,\kms\ from broadening. We find the metallicity is around Solar. 
These parameters are consistent with those in the literature (e.g. \citealt{1997A&AS..122...51K,2002A&A...393..897R,2011A&A...531A.143D,2015ApJ...804..146D}). 
HR\,5599 was claimed to be an A5m star by \citet{1968PASP...80..453C}, whilst \citet{1991A&AS...89..429R} listed it as an A4m star and noted its identification as a metallic-line star to be doubtful. 
We do not see evidence in the spectra that HR\,5599 is an Am star: the Ca K line is not weak and was fit with normal abundance for the star's \teff. 
\citet{1999mctd.book.....H} classified the star as A4--5IV--V . 
We found a spectral type of A6IV--V using the {\sc mkclass} spectral classification code of \citet{2014AJ....147...80G}. 

We calculated the distance to WASP-189 ($d = 99 \pm 1$\,pc) using the {\it Gaia} DR2 parallax of $9.999 \pm 0.075$\,mas \citep{2018A&A...616A...2L}, which we corrected to $10.08 \pm 0.11$\,mas using the systematic offset of $-82 \pm 33$\,$\mu$as suggested by \citet{2018ApJ...862...61S} to apply to bright, nearby stars. 
This is consistent with the {\sc Hipparcos} value of 10.29 $\pm$ 0.81\,mas \citep{2007A&A...474..653V}.  
We calculated the effective temperature ($T_{\rm eff,IRFM} = 7980 \pm 170$\,K) and angular diameter ($\theta = 0.218 \pm 0.011$\,mas) of the star using the infrared flux method (IRFM) of \citet{1977MNRAS.180..177B}, assuming reddening of $E(B-V) = 0.02$ estimated from the Na D lines.
We thus calculated its luminosity and its radius ($\log(L_*/L_{\rm \odot}) = 1.293 \pm 0.045$ and $R_{\rm *,IRFM} = 2.33 \pm 0.12$\,\rsol). 
We summarise the stellar properties in \cref{tab:stellar}. 

\begin{figure}
\includegraphics[width=90mm]{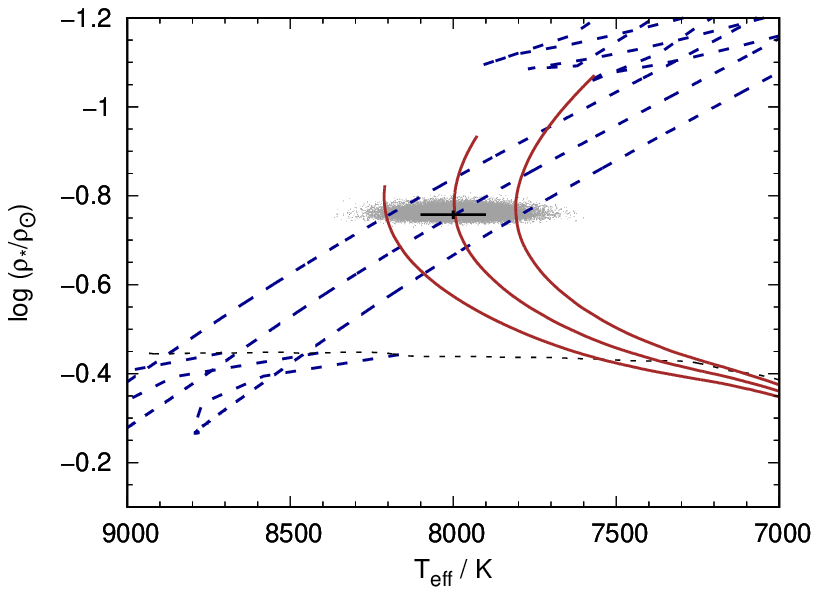}
\caption{
Modified Hertzsprung-Russell diagrams showing the results of the {\sc bagemass} MCMC analysis for WASP-189 using \rhostar\ from the combined analysis (\cref{sec:mcmc}) and \teff\ and [M/H] from \cref{tab:stellar}. 
The grey dots are the steps in the Markov chain. 
The dotted line (black) is the ZAMS. 
The solid lines (brown) are isochrones for $\tau = 0.855 \pm 0.080$\,Gyr . 
The dashed lines (blue) are evolutionary tracks for \mstar\ = 1.887 $\pm$ 0.057\,\msol. 
The black point with error bars are the values of \teff\ and \densstar\ measured from the spectra and the transit lightcurves, respectively.
\label{fig:evol}}
\end{figure}

We inferred \mstar\ = 1.887 $\pm$ 0.057\,\msol\ and age $\tau$ = 0.855 $\pm$ 0.080\,Gyr using the {\sc bagemass} stellar evolution MCMC code of \citet{2015A&A...575A..36M} with input of the values of \densstar\ from our adopted combined analysis (\cref{sec:mcmc}) and \teff\ and [Fe/H] from the spectral analysis (\cref{fig:evol}). 
We note that the mass and age are consistent with the values from \citet{2015ApJ...804..146D} (\mstar\ = 1.9 $\pm$ 0.2\,\msol; $\tau = 0.83 \pm 0.15$\,Gyr), and the mass is within 1\,$\sigma$ of the adopted value from our combined analysis (\cref{sec:mcmc}). 
By combining our values of \mstar\ and $R_{\rm *,IRFM}$, we calculated \logg\ = 3.98 $\pm$ 0.06.

Our values of \teff\ and $L_*$ suggest that WASP-189 is in the instability strip and may be a $\delta$-Scuti star. If so then it may exhibit photometric variations with amplitude of a few mmag to tenths of a mag and with a period of 18\,min to 8\,hr \citep{2010aste.book.....A}. 
Similarly, our values for \teff\ and \logg\ (from both \cref{tab:stellar} and \cref{tab:mcmc}) place WASP-189 in the region occupied by delta-Scuti stars (see the bottom panel of fig.~4 in \citealt{2017MNRAS.465.2662S}). 
This may explain the correlated noise in our follow-up transit lightcurves (\cref{fig:rv-tran}).
We searched the WASP lightcurves for modulation using the method of \citet{2011PASP..123..547M}. We found no convincing signal and place an upper limit of $\sim$5\,mmag on the amplitude of any sinusoidal signal. Further observation is required to test whether the star is a delta Scuti, which will be done by TESS \citep{2015JATIS...1a4003R}.

\begin{table}
\small
\caption{Stellar parameters} 
\label{tab:stellar}
\begin{tabular}{lccc}
\hline
\hline
Parameter & Symbol & Value & Unit \\
\hline 
Designations & \ldots & WASP-189 & \ldots \\
 & \ldots & HR\,5599 & \ldots \\
 & \ldots & HD\,133112 & \ldots \\
 & \ldots & HIP\,73608 & \ldots \\
Constellation & \ldots & Libra & \ldots \\
Right Ascension & \ldots & $\rm 15^{h} 02^{m} 44\fs86$ & \ldots \\
Declination		& \ldots & $\rm -03\degr 01\arcmin 52\fs9$	& \ldots \\	
Tycho-2 $V_{\rm mag}$	& \ldots & 6.64	& \ldots \\
2MASS $K_{\rm mag}$	& \ldots & 6.06	& \ldots \\
Spectral type   & \ldots & A6IV--V & \ldots \\
Stellar eff. temperature & $T_{\rm eff}$ & 8000 $\pm$ 100 & K \\
Stellar mass & \mstar & 1.887 $\pm$ 0.057 & \msol \\
Stellar radius (IRFM) & $R_{\rm *,IRFM}$ & 2.33 $\pm$ 0.12 & \rsol \\
Stellar surface gravity & $\log g_{*}$ & 3.98 $\pm$ 0.05 & [cgs]  \\
Stellar metallicity & [M/H] & $\sim$Solar & \ldots\\
Stellar luminosity & $\log$(\mstar/\msol) & 1.293 $\pm$ 0.045 & \ldots\\
Proj. st. rot. velocity & $v \sin i_{\rm *,spec}$ & $100 \pm 5$ & \kms\\
Reddening & $E(B-V)$ & 0.02 & \ldots\\
Distance & d & $99 \pm 1$ & pc \\
Age & $\tau$ & $0.855 \pm 0.080$ & Gyr \\
\hline
\end{tabular}
\end{table}

\section{System parameters from MCMC analysis}
\label{sec:mcmc}

We determined the system parameters from a simultaneous fit to the lightcurves, the RVs, and the CCFs using the current version of the Markov-chain Monte Carlo (MCMC) code presented in \citet{2007MNRAS.380.1230C} and described further in \citet{2010MNRAS.403..151C} and \citet{2015A&A...575A..61A}. 
We modelled the transit lightcurves using the formulation of 
\citet{2002ApJ...580L.171M} and limb-darkening was accounted for using the four-parameter non-linear law of \citet{2000A&A...363.1081C} (see Table~\ref{tab:mcmc} for the interpolated coefficients). 
The fitted parameters were orbital period $P$, transit epoch $T_{\rm c}$, transit duration $T_{\rm 14}$, planet-to-star area ratio $R_{\rm P}^{2}$/R$_{*}^{2}$, and impact parameter $b$.
We parameterised the Keplerian RV orbit by $T_{\rm c}$, $P$, the stellar reflex velocity semi-amplitude $K_1$, and the systemic velocity $\gamma$ (one per dataset).
Following \citet{2012MNRAS.422.1988A}, we assumed a circular orbit. 

When fitting the Rossiter-McLaughlin (RM) effect (i.e. the apparent radial-velocity anomaly that occurs during transit; e.g. \citealt{2012ApJ...757...18A}) we modelled it using the formulation of \citet{2011ApJ...742...69H}. 
This was parameterised by \svsicos\ and \svsicos, where $\lambda$ is the projected stellar obliquity.
When we instead modelled the planet's Doppler shadow in the average stellar line profiles (CCFs) during transit we employed the method presented in \citet{2010MNRAS.403..151C} and used in \citet{2010MNRAS.407..507C}, \citet{2012ApJ...760..139B,2017MNRAS.464..810B} and \citet{2017MNRAS.471.2743T,2018MNRAS.480.5307T}. 
The fitted parameters were $b$, the projected stellar rotation speed \vsinidt, $\lambda_{\rm DT}$, the FWHM of the line-profile perturbation due to the planet $v_{\rm FWHM}$, and the systemic velocity $\gamma_{\rm DT}$. 

Though we can measure stellar density, \densstar, directly from the transit lightcurves, we require a constraint on stellar mass \mstar, or radius \rstar, for a full characterisation of the system. 
For this we used our parallax-derived radius to place a Gaussian prior: $R_{\rm *,IRFM}$ = 2.33 $\pm$ 0.12 \rsol.

We performed three seperate MCMC analyses to obtain seperate measurements of the stellar obliquity: the first analysis (\mcmcrm) included the transit RVs from both spectrographs, the second analysis (\mcmch) included the HARPS transit CCFs, and the third analysis (\mcmcc) included the CORALIE transit CCFs. 
As we did not fit the RM effect in the latter two analyses, we excluded the RVs taken on the transit night, except for the final four taken by HARPS and the first four taken by CORALIE (all were outside of transit). Otherwise, all RVs were included in each analysis.

We accounted for stellar noise in the RV measurements by adding in quadrature with the formal RV uncertainties the level of `jitter' required to achieve $\chi^2_{\rm reduced} = 1$. 
The jitter values in \mcmch\ and \mcmcc\ were: 48\,\ms\ (HARPS orbit) and 74\,\ms\ (CORALIE orbit).
The jitter values in \mcmcrm\ were: 58\,\ms\ (HARPS orbit), 35\,\ms\ (CORALIE orbit), 35\,\ms\ (HARPS transit), and 95\,\ms\ (CORALIE transit).
To account for instrumental and astrophysical offsets, we partitioned the RV datasets and fit a seperate systemic velocity to each of them. 

\begin{figure}
\includegraphics[width=90mm]{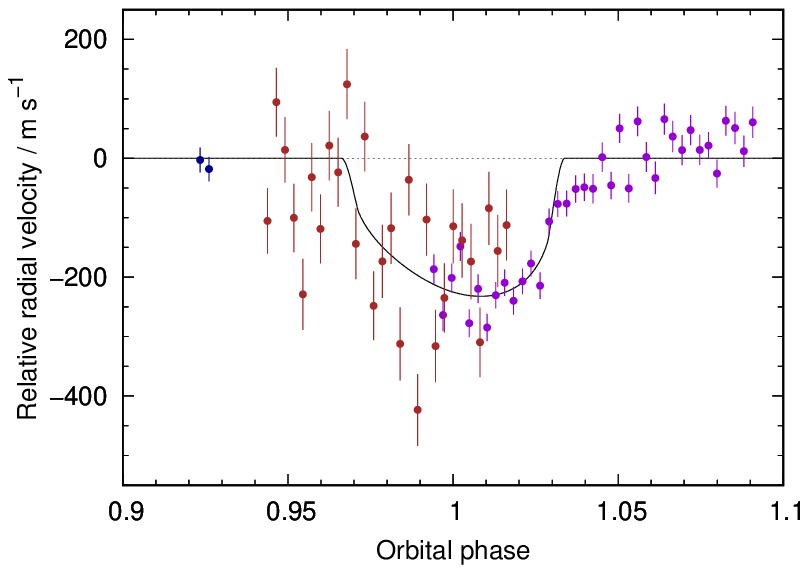}
\caption{The RM effect of WASP-189b revealed by RVs from CORALIE (brown) and HARPS (violet). The best-fitting orbital model has been subtracted. The best-fitting RM model is plotted. The RM effect produces an apparent blue-shift that varies little between the second and third transit contacts, indicating that the planet is in a polar orbit crossing the red-shifted hemisphere of the star. 
\label{fig:rm}}
\end{figure}

\begin{figure*}
\includegraphics[]{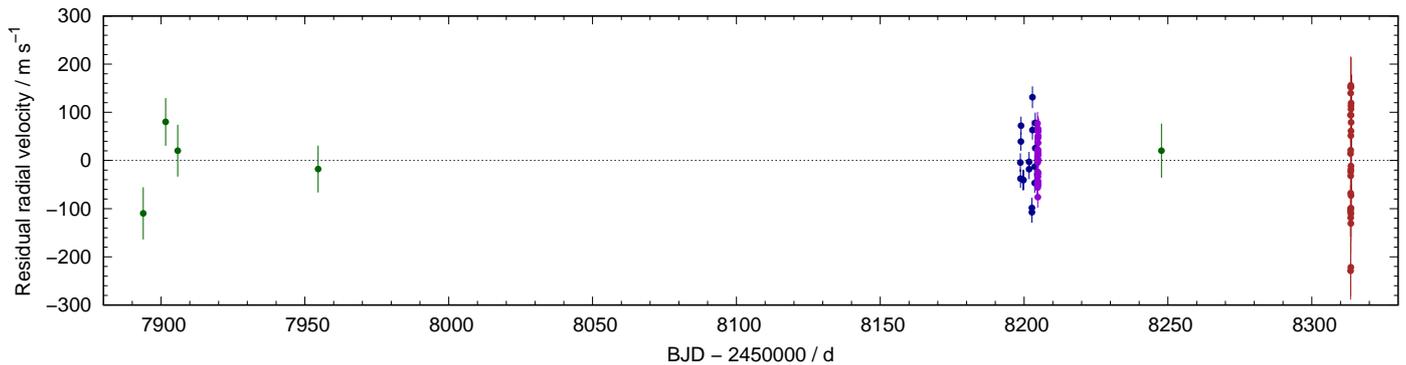}
\caption{
The residual RVs about the best-fitting Keplerian orbital and RM effect models from the \mcmcrm\ analysis (CORALIE orbital RVs = green; CORALIE RM RVs = brown; HARPS orbital RVs = blue; HARPS RM RVs = violet). The data are consistent with no variation, though only the the CORALIE orbital RVs span a significant duration.
\label{fig:rv-resid}}
\end{figure*}

We present in \cref{tab:mcmc} the results of the three analyses, which are fully consistent with each other. 
We plot the Doppler tomograms (i.e. the residual maps of the CCF time-series) both before and after removal of the planet model in \cref{fig:tomog}, and the fit to the RM effect in \cref{fig:rm}. 
We plot the fits to the transit lightcurves and the RVs in \cref{fig:rv-tran} for our adopted solution (\mcmch, as justified below).
We see no variation in the full RV residuals (\cref{fig:rv-resid}), as may be caused by an additional body, though the constraint is weak. 

The main point to note in comparing the three solutions is that the impact parameter (and therefore related parameters) is far better constrained in the tomographic analyses than in the RM-effect analysis. 
The impact parameter is relatively poorly constrained by the available transit photometry, but it is well constrained by the path that the planet traces out in the CCF time-series. 
We adopt the \mcmch\ solution due to the high S/N of the HARPS CCFs, which results in a higher precision for $b$ and related parameters. 

We find a projected stellar obliquity of $\lambda = 89.3 \pm 1.4^\circ$ (projected because the inclination of the stellar spin axis relative to the line of sight $i_*$ is unknown). 
Following the method of \citet{2011Ap&SS.331..485I}, we can place limits on the true stellar obliquity $\Psi$ by requiring the stellar rotation speed to be smaller than the break-up speed. For our adopted solution, we obtain 1-$\sigma$ limits of: $11.7^\circ < i* < 168.3^\circ$ and $\Psi = 90.0 \pm 5.8^\circ$. 
We can place such a tight constraint on $\Psi$ as it depends little on $i_*$ for polar orbits. 

\begin{table*} 
\small
\centering
\caption{System parameters} 
\label{tab:mcmc}
\begin{tabular}{lccccc}
\hline
Parameter & Symbol & \mcmcrm & \mcmch & \mcmcc & Unit\\ 
			&		&		& (Adopted solution)	&	&	\\
\hline 
\\
\multicolumn{5}{l}{\it MCMC Gaussian priors}\\
Stellar radius & $R_{\rm *}$ & 2.33 $\pm$ 0.12& 2.33 $\pm$ 0.12 & 2.33 $\pm$ 0.12 & $R_{\rm \odot}$ \\
Stellar effective temperature & $T_{\rm eff}$ & 8000 $\pm$ 100 & 8000 $\pm$ 100 & 8000 $\pm$ 100 & K \\
\\
\multicolumn{5}{l}{\it MCMC parameters controlled by Gaussian priors}\\
Stellar radius & $R_{\rm *}$ & 2.33 $\pm$ 0.12 & 2.33 $\pm$ 0.12 & $2.33 \pm 0.12$ & $M_{\rm \odot}$\\
Stellar effective temperature & $T_{\rm eff}$ & $8015 \pm 94$ & $7996 \pm 99$ & $8000 \pm 100$ & K \\
\\
\multicolumn{5}{l}{\it MCMC fitted parameters}\\
Orbital period & $P$ & $2.7240338 \pm 0.0000067$ & $2.7240330 \pm 0.0000042$ & $2.7240317 \pm 0.0000043$ & d\\
Transit epoch (HJD) & $T_{\rm c}$ & $2456706.4558 \pm 0.0023$ & $2456706.4545 \pm 0.0018$ & $2456706.4543 \pm 0.0016$ & d\\
Transit duration & $T_{\rm 14}$ & $0.1819 \pm 0.0065$ & $0.1813 \pm 0.0011$ & $0.1796 \pm 0.0034$ & d\\
Planet-to-star area ratio & $R_{\rm P}^{2}$/R$_{*}^{2}$ & $0.00372 \pm 0.00038$ & $0.00368 \pm 0.00026$ & $0.00367 \pm 0.00030$ & \ldots \\
Impact parameter$^{\rm a}$ & $b$ & $0.51 \pm 0.24$ & $0.4537 \pm 0.0072$ & $0.423 \pm 0.017$ & \ldots \\
Reflex velocity semi-amplitude & $K_{\rm 1}$ & $175 \pm 15$ & $182 \pm 13$ & $182 \pm 15$ & \ms\\
Systemic velocity (HARPS orbit)$^{\rm b}$ & $\gamma_{\rm RV}$ & $-24\,462 \pm 15$ & $-24\,452 \pm 12$ & $-24\,452 \pm 12$ & m s$^{-1}$ \\
Systemic velocity (CORALIE orbit)$^{\rm b}$ & $\gamma_{\rm RV}$ & $-25\,149 \pm 28$ & $-25\,172 \pm 31$ & $-25\,172 \pm 31$ & m s$^{-1}$ \\
Systemic velocity (HARPS RM) & $\gamma_{\rm RV}$ & $-24\,451.9 \pm 7.0$ & \ldots & \ldots & m s$^{-1}$ \\
Systemic velocity (CORALIE RM) & $\gamma_{\rm RV}$ & $-25\,173 \pm 21$ & \ldots & \ldots & m s$^{-1}$ \\
Systemic velocity (Tomography) & $\gamma_{\rm DT}$ & \ldots & $-25\,257.9 \pm 5.0$ & $-25\,241.6 \pm 1.2$ & m s$^{-1}$ \\
Orbital eccentricity & $e$ & 0 (assumed) & 0 (assumed) & 0 (assumed) & \ldots \\
Projected stellar rotation speed & $v \sin i_{\rm *,DT}$ & $108^{+ 96}_{- 38}$ & $97.1 \pm 2.1$ & $100.19 \pm 0.10$ & \kms\\
Intrinsic linewidth & $v_{\rm FWHM}$ & \ldots & $11.50 \pm 0.29$ & $16.5 \pm 3.6$ & \kms\\
Projected stellar obliquity & $\lambda$ & $83.3^{+4.7}_{-7.9}$ & $89.3 \pm 1.4$ & $88.0 \pm 1.5$ & $^\circ$ \\
\\
\multicolumn{5}{l}{\it MCMC derived parameters}\\
Stellar obliquity & $\Psi$ & $93.7 \pm 7.3$ & $90.0 \pm 5.8$ & $89.9 \pm 5.4$ & $^\circ$ \\
Scaled semi-major axis & $a/R_{\rm *}$ & $4.4 \pm 0.8$ & $4.591 \pm 0.041$ & $4.689 \pm 0.086$ & \ldots \\
Orbital inclination & $i$ & $83.4 \pm 4.7$ & $84.321 \pm 0.097$ & $84.81 \pm 0.31$ & $^\circ$\\
Ingress and egress duration & $T_{\rm 12}=T_{\rm 34}$ & $0.0138 \pm 0.0073$ & $0.01289 \pm 0.00043$ & $0.01239 \pm 0.00052$ & d\\
Stellar mass & $M_{\rm *}$ & $2.0 \pm 1.0$ & $2.20 \pm 0.34$ & $2.36 \pm 0.39$ & $M_{\rm \odot}$\\
Stellar surface gravity & $\log g_{*}$ & $4.00 \pm 0.25$ & $4.046 \pm 0.024$ & $4.076 \pm 0.033$ & [cgs]\\
Stellar density & $\rho_{\rm *}$ & $0.158 \pm 0.064$ & $0.1748 \pm 0.0037$ & $0.186 \pm 0.011$ & $\rho_{\rm \odot}$\\
Planetary mass & $M_{\rm P}$ & $1.86 \pm 0.66$ & $2.13 \pm 0.28$ & $2.21 \pm 0.30$ & $M_{\rm Jup}$\\
Planetary radius & $R_{\rm P}$ & $1.38 \pm 0.10$ & $1.374 \pm 0.082$ & $1.370 \pm 0.090$ & $R_{\rm Jup}$\\
Planetary surface gravity & $\log g_{\rm P}$ & $3.36 \pm 0.31$ & $3.414 \pm 0.040$ & $3.438 \pm 0.051$ & [cgs]\\
Planetary density & $\rho_{\rm P}$ & $0.73 \pm 0.31$ & $0.83 \pm 0.10$ & $0.88 \pm 0.14$ & $\rho_{\rm Jup}$\\
Orbital semi-major axis & $a$ & $0.04770 \pm 0.0089$ & $0.0497 \pm 0.0026$ & $0.0508 \pm 0.0028$ & AU\\
Planetary equilibrium temperature:$^{\rm c}$  \\
\hspace{3mm} full redistribution & $T_{{\rm eql},f=1}$ & $2690 \pm 260$ & $2641 \pm 34$ & $2611 \pm 41$ & K\\
\hspace{3mm} dayside redistribution & $T_{{\rm eql},f=2}$ & $3200 \pm 320$ & $3140 \pm 41$ & $3105 \pm 49$ & K\\
\hspace{3mm} instant re-radiation & $T_{{\rm eql},f=8/3}$ & $3440 \pm 340$ & $3374 \pm 44$ & $3337 \pm 53$ & K\\
\\
\multicolumn{5}{l}{\it Limb-darkening coefficients}\\
c1 & \ldots & 0.469		& 0.465		& 0.466		& \ldots\\
c2 & \ldots & 0.431		& 0.457		& 0.448		& \ldots\\
c3 & \ldots & $-$0.357	& $-$0.394	& $-$0.380	& \ldots\\
c4 & \ldots & 0.093		& 0.110		& 0.103		& \ldots\\
\\ 
\hline 
\end{tabular} 
\raggedright
{\it Note:} $^a$ Impact parameter is the distance between the centre of the stellar disc and the transit chord: $b = a \cos i / R_{\rm *}$.
\newline $^b$ Note that the orbital RV datasets were identical for \mcmch\ and \mcmcc, but each contained four fewer points for \mcmcrm.
\newline $^c$ \teql\ was calculated assuming a zero-albedo planet using $T_{\rm eql} = f^{1/4}T_{\rm eff}\sqrt{R_{\rm *}/{2a}}$, where $f$ parameterises the redistribution of heat. The instant re-radiation case is dealt with by \citet{2011ApJ...729...54C}.
\end{table*}

\section{Discussion}
We have presented the discovery of WASP-189b, an ultra-hot Jupiter (\mplanet\ = 2.13 $\pm$ 0.28\,\mjup; \rplanet\ = 1.374 $\pm$ 0.082\,\rjup) in a 2.72-d, polar orbit ($\Psi = 90.0 \pm 5.8^\circ$; $\lambda = 89.3 \pm 1.4^\circ$) around the rapidly rotating A6IV--V star WASP-189 (HR\,5599).
The predicted equilibrium temperature of WASP-189b (\teql\ = 2641 $\pm$ 34\,K; assuming zero albedo and efficient heat redistribution) is the third highest for the known exoplanets, behind KELT-9b (\teql\ = 4050 $\pm$ 180\,K; \citealt{2017Natur.546..514G}) and WASP-33b (\teql\ = 2710 $\pm$ 50\,K; \citealt{2010MNRAS.407..507C}). 
Our values of \teff\ and $L_*$ suggest that WASP-189 is in the instability strip and may be a $\delta$-Scuti star like WASP-33 \citep{2010MNRAS.407..507C, 2011A&A...526L..10H}.

WASP-189b is set to be engulfed by its star as it ascends the giant branch: the star is predicted to expand to the planet's current orbit in $\sim$320\,Myr (e.g. \citealt{2017ApJ...838..161S}). 
As was noted by \citet{2017Natur.546..514G} for KELT-9, this may result in a bright transient event \citep{2012MNRAS.425.2778M} and an anomalously rapidly rotating red giant enriched in lithium \citep{2016ApJ...829..127A}. 

WASP-189 ($V = 6.64$) is the brightest known host of a transiting hot Jupiter by almost a magnitude.\footnote{There are six known hosts of non-transiting hot Jupiters brighter than WASP-189 in the range $V = 4.1$--$6.3$: ups And, tau Boo, 51\,Peg, HD\,217107, HD\,179949 and HD\,75289.} The next brightest are the A stars KELT-9 ($V=7.56$) and KELT-20/MASCARA-2 ($V=7.59$; \citealt{2017AJ....154..194L,2018A&A...612A..57T}), and the well-studied later-type stars HD\,209458 ($V=7.65$; G0V; \citealt{2000ApJ...529L..45C,2000ApJ...529L..41H}) and HD\,189733 ($V=7.68$; K1--2V; \citealt{2005A&A...444L..15B}). 
WASP-189 is the third-brightest star known to host a transiting exoplanet of any mass. The two brighter stars host multiple planets and transiting super-Earths: HD\,219134 ($V=5.57$; \citealt{2015A&A...584A..72M, 2017NatAs...1E..56G}) and 55\,Cnc ($V=5.95$; \citealt{2004ApJ...614L..81M,2018arXiv180704301B}).

The transits of WASP-189b are among the shallowest detected by WASP and confirmed to be caused by an orbiting substellar body, 
with a transit depth (4\,mmag) similar to WASP-30b, WASP-71b, WASP-72b, WASP-73b, WASP-99b, and WASP-136b
\citep{2011ApJ...726L..19A, 2013A&A...552A.120S, 2013A&A...552A..82G, 2014A&A...563A.143D, 2014MNRAS.440.1982H, 2017A&A...599A...3L}. 

\begin{figure}
\includegraphics[width=90mm]{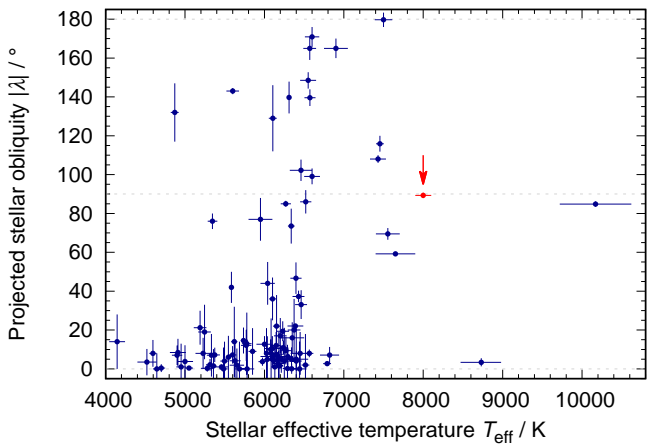}
\caption{The distribution of projected stellar obliquity $\lambda$ with stellar effective temperature \teff, excluding systems with $\sigma_{\lambda} > 20^\circ$. WASP-189 is the red point indicated by an arrow. Of those systems with \teff\ $>$7000\,K, the orbit of just one planet may be aligned with its star's spin. The other seven systems are misaligned: one orbit is retrograde, two orbits are polar, and the remaining four orbits are evenly split 20--30$^\circ$ either side of polar. 
\label{fig:lambda-teff}}
\end{figure}

With the most confidently polar orbit ($\Psi = 90.0 \pm 5.8^\circ$) of any system, WASP-189 continues the trend that the spins of hot stars tend to be misaligned with the orbits of their hot Jupiters (\cref{fig:lambda-teff}; \citealt{2010ApJ...718L.145W}). 
We note that only one of the eight known systems with \teff\ $>$ 7000\,K is aligned (KELT-20/MASCARA-2). Of the three systems with \teff\ $\geq$ 8000\,K, one is aligned and two have polar orbits (WASP-189 and KELT-9). 
Perhaps this is telling us something about the states that systems are driven toward by tides \citep{2012MNRAS.423..486L,2013ApJ...769L..10R}, angular momentum transport by internal gravity waves \citep{2012ApJ...758L...6R}, or a combination of the two. 

\section*{Acknowledgements}
WASP-South is hosted by the South African Astronomical Observatory; we are grateful for their ongoing support and assistance. 
Funding for WASP comes from consortium universities and from the UK's Science and Technology Facilities Council. 
The Swiss {\it Euler} Telescope is operated by the University of Geneva, and is funded by the Swiss National Science Foundation. 
The research leading to these results has received funding from the  ARC grant for Concerted Research Actions, financed by the Wallonia-Brussels Federation. TRAPPIST is funded by the Belgian Fund for Scientific Research (Fond National de la Recherche Scientifique, FNRS) under the grant FRFC 2.5.594.09.F, with the participation of the Swiss National Science Fundation (SNF). MG and EJ are FNRS Senior Research Associates.
Based on observations collected at the European Organisation for Astronomical Research in the Southern Hemisphere under ESO programme 0100.C-0847.
This research has made use of TEPCat, a catalogue of the physical properties of transiting planetary systems maintained by John Southworth.





%







\bsp	
\label{lastpage}
\end{document}